\documentstyle[12pt,psfig]{article}
\begin{document}

{\flushright{\small NSF-ITP/97-146\\NUHEP-TH-97-14\\
astro-ph/9711288\\}}

\vspace{0.2in}
\begin{center}
{\Large  {\bf Dark Matter with Time-Dependent Mass}}\footnote{Based
on a talk by SMC at COSMO-97, International Workshop on 
Particle Physics and the Early Universe,
15-19 September 1997, Ambleside, Lake District, England.}

\vspace{0.2in}
Greg W. Anderson
\vskip 0.2cm
{{\it Dept. of Physics and Astronomy, Northwestern University
\\ 2145 Sheridan Rd., Evanston, IL 60208-3112, USA}
\\
\small Email: {\tt ganderson@nwu.edu}}

\vspace{0.2in}
Sean M. Carroll
\vskip 0.2cm
{{\it Institute for Theoretical Physics, University of
California, \\
Santa Barbara, California 93106, USA}
\\
\small Email: {\tt carroll@itp.ucsb.edu}}
\end{center}

\vskip 1truecm

\begin{abstract}

We propose a simple model in which the cosmological dark matter 
consists of particles whose mass increases with the scale factor of
the universe.  The particle mass is generated by the expectation
value of a scalar field which does not have a stable vacuum state, 
but which is effectively stabilized by the rest energy of the ambient
particles. As the universe expands, the density of particles decreases, 
leading to an increase in the vacuum expectation value of the scalar (and 
hence the mass of the particle).  The energy density of the coupled system
of variable-mass particles (``vamps'') redshifts more slowly 
than that of ordinary matter.  Consequently, the age of the universe 
is larger than in conventional scenarios.
\end{abstract}

\medskip

\newpage

\renewcommand{\baselinestretch}{1}
\baselineskip 18pt

\section{Introduction}
\label{sec:introduction}

The Big Bang model has proven extraordinarily successful as a
framework for interpreting the structure and evolution of the
universe on large scales.  Within that framework, the cold dark
matter scenario (featuring massive particles which bring the 
density of the universe to its critical value, and a scale-free
spectrum of Gaussian density perturbations) has provided an
elegant theory of structure formation, which unfortunately
seems to fall short of perfect agreement with observation.  
Although the precise extent to which CDM disagrees with
observation is arguable, there are two important areas in which
the discrepancies are particularly troubling: 
predicting an age for the universe which is larger than the ages
of the oldest globular clusters, and matching the
COBE-normalized power spectrum of density fluctuations as measured
by microwave background anisotropy experiments and direct studies
of large-scale structure.

One way in which the simple CDM scenario may be modified, affecting
the age of the universe as well as the evolution of density fluctuations,
is to imagine that the closure density is provided by something
different than (or in addition to) nonrelativistic particles.
In a flat Robertson-Walker universe with metric
\begin{equation}
  ds^2 = -dt^2 + a^2(t)(dx^2 + dy^2 +dz^2)
  \label{metric}
\end{equation}
and energy-momentum tensor
\begin{equation}
  T^\mu{}_\nu = {\rm diag}(-\rho,p,p,p)\ ,
\end{equation}
the Friedmann equations imply that the time derivative of the scale
factor $a$ satisfies
\begin{equation}
  \dot a^2 = {{8\pi G}\over{3}}a^2\rho\ .
\end{equation}
The evolution of $\dot a$ is therefore dependent on how the energy
density $\rho$ scales with $a$; if $\rho \propto a^{-n}$, we have
\begin{equation}
  {{\ddot a}\over{a}} = {{4\pi G}\over{3}}(2-n)\rho\ .
  \label{ddota}
\end{equation}
Hence, the more slowly the energy density decreases
as the universe expands, the more slowly the expansion will decelerate,
implying a correspondingly older universe for any given value of the
expansion rate today --- for a flat universe dominated by such an
energy density, the age is $t_0 = {2\over n}H_0^{-1}$,
where $H={\dot a}/a$ is the Hubble parameter and the subscript $0$
refers to the present time.
(Eq.~(\ref{ddota}) can also be derived by positing an equation
of state $p=w\rho$ and using energy conservation; the two 
parameterizations are related by $n=3(1+w)$.)
The energy density in a species of ordinary 
``matter'' (a nonrelativistic particle species $X$) can be written 
$\rho_X = m_X n_X$, where $m_X$ is the mass of the particle and 
$n_X$ is its number density.  The energy density of a matter-dominated
universe is therefore proportional to $a^{-3}$, as the mass stays
constant while the number density is proportional to the volume;
the age of such a universe is $t_0 = {2\over 3}H_0^{-1}$.

Although there is some controversy over the value of the Hubble
constant, most recent determinations are consistent with a value
$H_0 = 70\pm 10{\rm ~km/sec/Mpc}$, or $H_0^{-1} = 
(14\pm 2)\times 10^9{\rm ~yr}$ \cite{h0}.  
The upper limit on the age of a matter-dominated flat
universe is therefore $t_0({\rm MD}) \leq 11\times 10^9{\rm ~yr}$.
Meanwhile, calculations of the ages of globular clusters imply an
age $t_{\rm GC} \sim 15\times 10^9{\rm ~yr}$, with a
lower bound of $t_{\rm GC} \geq 12\times 10^9{\rm ~yr}$ \cite{gcage}.
The apparent discrepancy between these values may be resolved
by a revision in distance determinations to globular clusters, as suggested
by recent measurements by the Hipparcos satellite \cite{hipp};
while this would be the simplest solution, further work is necessary
to accept it with confidence.

Alternative resolutions are provided by models
in which the density parameter $\Omega=1$, and some or all of the
unseen energy density resides in a component which redshifts more
slowly than nonrelativistic matter.  The most popular such alternative
is the introduction of a cosmological constant $\Lambda$, for which
$\rho_\Lambda ={\rm constant}$.  Such models have some attractive 
features, but are also plagued with both theoretical and observational 
disadvantages \cite{lambda,cpt}.  A popular variation
on this theme is to invoke a slowly-rolling scalar field, or
equivalently a cosmological constant whose value varies with time, or
simply an unspecified smooth component \cite{w}.
More speculative possibilities include a 
network of cosmic strings \cite{strings} or stable textures 
\cite{textures}.  We will not enumerate the
good and bad qualities of each of these scenarios, noting only that
none are sufficiently compelling to discourage the exploration of
still further models.

In this paper we propose a simple model in which the dark matter consists
of particles $\psi$ whose rest mass increases with time.  This is
achieved by having the rest mass derive from the expectation value of
a scalar field $\phi$; the potential for $\phi$ depends on the number
density of $\psi$ particles, and therefore increases naturally on 
cosmological timescales as the universe expands.  As a result, the particle
energy density $\rho_\psi = m_\psi n_\psi$ decreases more slowly than
$a^{-3}$, resulting in a larger age for the universe.  (There is also
a contribution from the potential energy of $\phi$, which redshifts at
the same rate.)  We discuss some of the cosmological consequences of
this proposal, including potential observational tests.  The question
of structure formation in the presence of such particles, as well as
the construction of realistic particle physics models containing the
necessary fields, is left for future work.

After this paper was first submitted, we became aware of earlier
an proposal for dark matter with time-dependent mass by Casas,
Garc\`ia-Bellido, and Quir\'os \cite{cgq}.  These authors considered
models of scalar-tensor gravity, in which the scalar coupled 
differently to different species of particles.

\section{Scale factor and age of the universe}
\label{sec:age}

The model consists of a scalar $\phi$ and a particle species $\psi$,
which can be either bosonic or fermionic for the purposes of this work.
The mass of $\psi$ is imagined to come from the vacuum expectation
value of $\phi$, with the constant of proportionality some dimensionless
parameter $\lambda$:
\begin{equation}
  m_\psi = \lambda \langle \phi \rangle\ .
\end{equation}
More elaborate dependences of $m_\psi$ on $\langle \phi \rangle$ are
certainly conceivable, but for the purposes of this paper we make
this simple choice.
The dynamics of $\phi$ are determined by a conventional kinetic term
and a potential energy $U(\phi)$. The notable feature of the model
is that we choose the potential $U(\phi)$ to blow up at $\phi=0$ and 
roll monotonically to zero as $\phi\rightarrow\infty$.  For simplicity 
we will write
\begin{equation}
  U(\phi)=u_o\phi^{-p}\ ,
  \label{potential}
\end{equation}
although more complicated forms are again possible.  While such a potential
seems unusual, this form can arise for example due to nonperturbative
effects lifting flat directions in supersymmetric gauge theories 
\cite{ads}, as well as for moduli fields in string theory.  
(In fact this form of potential is not strictly necessary,
as the phenomenon we will describe can occur with almost any potential;
however, the effects are most dramatic with this choice.)

This model possesses no stable vacuum state; in empty space $\phi$ tends 
to roll to infinity.  We consider instead the behavior of $\phi$ in
a homogeneous background of $\psi$'s with number density $n_\psi$.
In that case, the dependence of the free energy on the value of $\phi$
comes both from the potential $U(\phi)$ and the rest energy of the
$\psi$ particles, which have a mass proportional to $\phi$.  The equilibrium
value of a homogeneous configuration is therefore one which minimizes
an effective potential of the form
\begin{equation}
  V(\phi)=u_0\phi^{-p}+\lambda n_\psi \phi\ .
\end{equation}
(See Fig.~1.) 
\begin{figure}
  \vskip -3.75cm
  \centerline{
  \psfig{figure=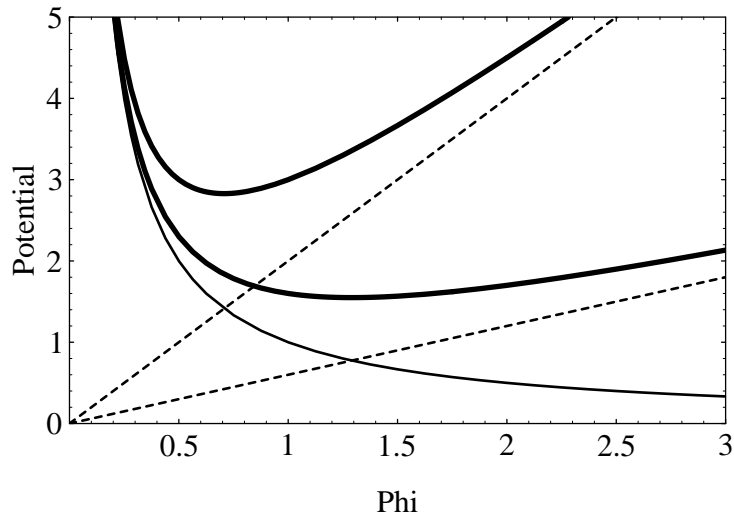,angle=0,height=13.5cm}}
  \vskip -3cm
  \caption{Effective potential for $\phi$.  The light solid curve is
  the bare potential $U(\phi)\propto \phi^{-1}$.  The effective potential
  at finite density, given by the solid curves, is obtained by adding
  a contribution linear in $\phi$ and proportional to the  
  number density $n_\psi$.  This is plotted for two different values of
  $n_\psi$, corresponding to two different stages in the evolution of
  the universe.  As the universe expands, $n_\psi$ decreases, and the
  equilibrium value of $\phi$ increases.}
\end{figure}
The additional contribution can be thought of as arising because 
increasing $\phi$ increases the energy density in $\psi$'s since
it increases the mass of $\psi$.
The expectation value of $\phi$ is therefore
\begin{equation}
  \langle\phi\rangle=\left({{pu_0}\over{\lambda n_\psi}}
  \right)^{1/(1+p)}\ .
  \label{expphi}
\end{equation}
(Such a configuration is not truly stable, as spatially inhomogeneous
perturbations will tend to grow, but it can be stable enough for
cosmological purposes.)  Density-dependent potentials such as
this have been discussed previously in other contexts; see
{\it e.g.} \cite{ddp}.

In an expanding universe, the number density $n_\psi$ will change
with time;
in turn, the mass of both $\phi$ and $\psi$ will change, as 
will the vacuum energy. After the interactions of $\psi$ have frozen
out, the number density can be written as $n_\psi = n_{\psi 0}a^{-3}$,
where $n_{\psi 0}$ is the density when $a=1$, which we take to be
the present epoch.  Then $\phi$ evolves as
\begin{equation}
  \phi=\phi_0 a^{3/(1+p)}\ , 
\end{equation}
where $\phi_0$ is the value of (\ref{expphi}) at the present time.
In terms of these variables the mass of the $\phi$ boson
is given by
\begin{equation}  m^2_\phi={{\partial^2V}\over{\partial\phi^2}}
  =\left[p(p+1)u_0\phi_0^{-(p+2)}\right]a^{-3(2+p)/(1+p)}\ ,
\end{equation}
and the mass of $\psi$ is
\begin{equation}
  m_\psi = \lambda \phi_0 a^{3/(1+p)}\ .
\end{equation} 
Both $\psi$ and $\phi$ are therefore variable-mass particles, or
``vamps''; a cosmological model in which vamps are the dominant
component of the energy density at late times will be referred to
as VDM.

There are a number of contributions to the energy density of the
universe in this model.  These include the energy in the scalar
$\phi$ particles, in the $\psi$ particles, in the time derivative
of the expectation value of $\phi$, in the potential $U(\phi)$, 
and in ordinary components of matter and radiation.  For reasonable
values of the parameters, the energies in $\dot\phi^2$ and in
$\phi$ quanta are small; the former because $\phi$ is only changing
on cosmological timescales, and the latter because the mass of
$\phi$ is decreasing with time.  (At early times, the $\phi$ bosons
are very massive and rapidly decay.)
The important new contribution is therefore
simply $V(\phi)$, the sum of the fundamental potential and the
rest energy in the $\psi$'s.  (We assume for now that $\psi$ is
is nonrelativistic.  As we discuss later, it is most likely that the
particles were relativistic when they decoupled, but at late times
their momenta have redshifted sufficiently that they are slowly moving
today.)  Both of these components turn out to depend on the scale 
factor in the same way; the ratio of the energy density in $\psi$
particles to that in the potential for $\phi$ is simply
\begin{equation}
  {{\rho_\psi}\over{\rho_{U(\phi)}}} = {1\over p}\ .
\end{equation}
It is therefore convenient to deal with the sum of these two 
contributions,
\begin{equation}
  \rho_{V} = (1+p)u_0 \phi^{-p} = 
  \left({{1+p}\over p}\right)\lambda \phi n_{\psi}\ ,
\end{equation}
which evolves as
\begin{equation}
  \rho_V = \rho_{V0} a^{-3p/(1+p)}\ .
\end{equation}
The parameter $w$ characterizing the effective equation of state of the
coupled $\phi$/$\psi$ system is therefore $w=-1/(1+p)$.

The energy density $\rho_M$ in ordinary massive particles
(baryons plus a possible cold dark matter component) redshifts
as $a^{-3}$, more slowly than $\rho_V$, and will therefore be
the dominant source of energy density in the universe for 
intermediate redshifts.  
The redshift at which $\rho_V = \rho_M$ is given by
\begin{equation}
  1+z_{VM}= \left({{\rho_{V0}}\over{\rho_{M0}}}
  \right)^{(1+p)/3}\ .
\end{equation}
The age of the universe, meanwhile,
will be larger than in conventional flat models.  The age
corresponding to a redshift $z$ is given by
\begin{equation}  \begin{array}{rcl}
   t &=& \displaystyle\int_0^{a} {{da'}\over{\dot a'}}\\
   &=& \displaystyle H_0^{-1}\int_0^{(1+z)^{-1}} \left[1-\Omega_0 
   +\Omega_{M0}x^{-1}
   +\Omega_{V0}x^{(2-p)/(1+p)}\right]^{-1/2}\, dx\ .
  \end{array}
\end{equation}
For the limiting case $\Omega_{M0}=0$, $\Omega_0=\Omega_{V0}=1$, 
we find that the age of the universe now is simply
\begin{equation}
  t_0={2\over 3}H_0^{-1}\left(1+ p^{-1}\right)\ .
\end{equation}
Fig.~2 plots the age of universes with 
$\Omega_0=1$ as a function of $\Omega_{M0}=1-\Omega_{V0}$, for $p=1$.  

An interesting feature of this model, in comparison with alternative
theories of rolling scalar fields and unusual equations of state,
is that (because $\psi$ is massive and nonrelativistic at late times)
it is at least conceivable that the energy density of the universe
is dominated solely by baryons and vamps (without any significant
cold dark matter component).
For illustrative purposes, let us define the ``minimal VDM model'' as
that with $p=1$, $\Omega_{V0} = 1-\Omega_{\rm baryon} = 0.96$,
and $H_0 = 70$~km/sec/Mpc.  This is a flat universe consisting
solely of baryons and vamps, with the baryon density consistent with
the prediction of Big Bang nucleosynthesis.  In this minimal model, 
vamp-matter equality occurs at a redshift $z_{VM}\sim 7$.  
The age of the universe turns out
to be approximately $17\times 10^9$~years, in good accord
with the (pre-Hipparcos) ages of the oldest globular clusters.

\begin{figure}
  \vskip -3.75cm
  \centerline{
  \psfig{figure=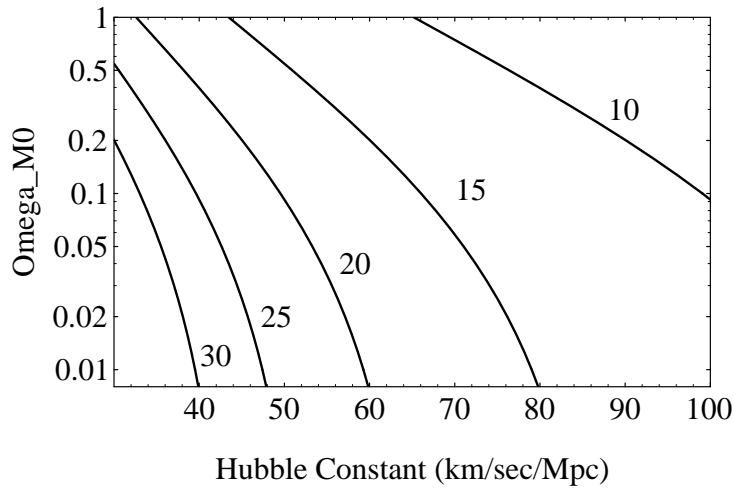,angle=0,height=13.5cm}}
  \vskip -3cm
  \caption{Age of the universe in billions of years.  The values in
  this plot are computed for flat universes consisting of only vamps
  and nonrelativistic matter, with $p=1$.}
\end{figure}

\section{Particle parameters and abundances}
\label{sec:abundances}

The properties we have deduced to this point depend on the present
energy density $\rho_{V0}$, but not on any assumptions about the
parameters of the  particle physics model in which we 
imagine the necessary fields and interactions
could arise. To understand the formation of 
large-scale structure in the model, however, it is necessary to know 
the mass and average velocity of the $\psi$ particles
today, and to compute these requires some detailed knowledge
of the interactions of our two fields.  In the absence of a specific
model, we will estimate these quantities under the minimal
assumptions that $\psi$ was in thermal equilibrium at some high
temperature and has evolved freely ever since.

We begin by considering the general problem of the motion of an
otherwise free particle whose mass $m_\psi=\lambda\phi$
may vary throughout spacetime.  The motion of such a particle
extremizes the action
\begin{equation}  \begin{array}{rcl}
  S &=& \displaystyle\int\sqrt{-p^\mu p_\mu}\, d\tau\\
  &=& \displaystyle\lambda\int \phi(x^\mu) \left(-g_{\mu\nu}
  {{dx^\mu}\over{d\tau}}{{dx^\nu}\over{d\tau}}\right)^{1/2}
  \, d\tau\ ,
  \end{array}
\end{equation}
where $\tau$ is the proper time along the particle's trajectory
and $p^\mu = m(dx^\mu/d\tau)$ is the particle's four-momentum.
Variation of this action with respect to the path leads to an
equation of motion
\begin{equation} 
  {{Dp^\mu}\over{d\tau}}\equiv
  {{dp^\mu}\over{d\tau}} + \Gamma^\mu_{\rho\sigma}
  {{dx^\rho}\over{d\tau}} p^\sigma = -\lambda \nabla^\mu\phi\ ,
  \label{geodesic1}
\end{equation}
which can be written explicitly in terms of the path $x^\mu(\tau)$ as
\begin{equation} 
  {{d^2x^\mu}\over{d\tau^2}} + \Gamma^\mu_{\rho\sigma}
  {{dx^\rho}\over{d\tau}} {{dx^\sigma}\over{d\tau}} =
  -\left(g^{\mu\nu} +{{dx^\mu}\over{d\tau}} {{dx^\nu}\over{d\tau}}
  \right)\partial_\nu(\ln{\phi})\ .
  \label{geodesic2}
\end{equation}
Since we are assuming that $\phi=\phi(t)$ is constant along 
spacelike hypersurfaces 
$t={\rm constant}$ of the metric (\ref{metric}), we can solve 
explicitly for the motion of a particle obeying (\ref{geodesic1}).
In terms of the magnitude of the spacelike 3-momentum,
\begin{equation} 
  |\vec{p}\, |^2 = g_{ij}p^i p^j = a^2 \delta_{ij} p^ip^j\ ,
  \label{momentum}
\end{equation}
we find
\begin{equation} 
  |\vec p\,| \propto a^{-1}\ ,
  \label{redshift}
\end{equation}
just as for conventional (constant-mass) particles.  The distinction
arises for the velocity; if the four-velocity satisfies $p^\mu = mu^\mu$,
the magnitude of the three-velocity $|\vec u|= (g_{ij}u^i u^j)^{1/2}$ is
proportional to $(a\phi)^{-1}$.  Thus, as the particles get more
massive with time, they naturally slow down even more rapidly than
ordinary test particles.

Although we have not specified any explicit interactions between
the vamps and visible matter, we may imagine 
that such reactions exist as long as they are sufficiently weak that
they do not lead to consequences which would have already been
observed.  As a result of such interactions, we presume that the
$\psi$'s were in thermal equilibrium at some high temperature.  Their
equilibrium phase-space distribution function is either a
Fermi-Dirac or Bose-Einstein distribution,
\begin{equation} 
  f(p) = {{g_\psi}\over{h_P^3}}{1\over{e^{E/kT_\psi}\pm 1}}\ ,
  \label{fermidirac}
\end{equation}
where $g_\psi$ is the number of spin degrees of freedom, $h_P$ is
Planck's constant, $k$ is Boltzmann's constant, $T_\psi$ is the 
temperature of the $\psi$'s, and $E$ is the energy, given by
\begin{equation} 
  E = p^0 = (m_\psi^2 + |\vec{p}\, |^2)^{1/2}\ .
  \label{energy}
\end{equation}
As the temperature and density increase, $|\vec{p}\,|$ goes up
while $m_\psi$ goes down.  At sufficiently early times, therefore,
the $\psi$ particles were relativistic, and $E \sim |\vec{p}\,|
\propto a^{-1}$.  Under these circumstances the $\psi$'s behave like 
ordinary relativistic particles; their temperature redshifts as $T_\psi
\propto a^{-1}$, and their energy density as $\rho_\psi \propto a^{-4}$.
When they become nonrelativistic, on the other hand, their kinetic
temperature will scale as $T_\psi \sim |\vec{p}\, |^2/m_\psi \propto 
a^{-2}\phi^{-1}$; they cool off more rapidly than ordinary matter.
(Strictly speaking it is incorrect to speak of a temperature after
the particles become nonrelativistic, as the varying rates at which
the particles slow down will distort the initially thermal distribution.)

It is reasonable to assume that $\psi$ was relativistic when it 
decoupled, and we may proceed under this assumption to show that it 
leads to a consistent picture.  In that case the number density
of $\psi$ today is given by the standard formula \cite{kt}
\begin{equation}
  n_{\psi 0} = 825\, r_\psi\ {\rm cm}^{-3} \ ,
\end{equation}
where $r_\psi$ is the ratio of $g_{\rm eff}$, the effective number of 
degrees of freedom in $\psi$, to $g_{*f}$, the total 
effective number of relativistic degrees of freedom at freeze-out.  
(In terms of the number of physical degrees of freedom $g$, 
$g_{\rm eff} = g$ for bosons and $g_{\rm eff} = 3g/4$ for fermions.)
For simplicity let us consider the case $p=1$.
Then we can directly determine the mass of $\psi$ in terms of
the current density parameter $\Omega_{\psi 0}$ and Hubble constant
$H_0 = 100h$~km/sec/Mpc:
\begin{equation}
  m_{\psi} = 12.7 \, \Omega_{\psi 0}h^2 r_\psi^{-1}\,a^{3/2}
  \ {\rm eV}\ .
  \label{mpsi}
\end{equation}
In terms of the Yukawa coupling $\lambda$, the other relevant
parameters of the model are then
\begin{equation}
  u_0 = 1.02\times 10^{-9} {{\Omega_{\psi 0}^2 h^4}\over
  {\lambda r_\psi}}
  \ ({\rm eV})^5
  \label{u0}
\end{equation}
and
\begin{equation}
  m_\phi = 1.00\times 10^{-6} \,{{\lambda r_\psi}\over
  {\Omega_{\psi 0}^{1/2} h}}\,a^{-9/4} \ {\rm eV}\ .
  \label{mphi}
\end{equation}
The temperature of the $\psi$ particles (while they are still relativistic)
is diluted somewhat with respect
to that of the photons, due to entropy production subsequent to
the freeze-out of $\psi$:
\begin{equation}
  \begin{array}{rcl}
  T_\psi &=& \left({{g_{*0}}\over{g_{*f}}}\right)^{1/3} 
  T_{\gamma 0} \,a^{-1}\\
  &=& 3.55\times 10^{-4}\,a^{-1}g_{*f}^{-1/3}
  \ {\rm eV}\ .
  \end{array}
  \label{tpsi} 
\end{equation}
Comparing (\ref{mpsi}) to (\ref{tpsi}), we find that the $\psi$'s first 
become non-relativistic at a redshift of 
\begin{equation}
  z_{NR}= 66.3\left({{\Omega_{\psi 0}h^2 g_{*f}^{1/3}}\over{r_\psi}}
  \right)^{2/5}\ .
\end{equation}

\section{Further consequences}
\label{sec:tests}

Although the VDM model helps to alleviate the age
problem, there are a number of other cosmological tests that
could conceivably rule it out.  For example, nucleosynthesis
places stringent limits on the number of degrees of freedom
contributing to the energy density at $T\sim 1{\rm ~keV}$ \cite{bbn}.  
Particles which decouple at sufficiently high energies are not
constrained by this test, as their number density is diluted by
entropy production after decoupling.  We do not know the temperature
at which $\psi$ decouples, although there is no reason to believe
that it isn't sufficiently high to evade the nucleosynthesis bound.
Meanwhile, the energy density in $\phi$ is much less than that 
in $\psi$ at high temperatures, and is therefore even less constrained.

Any model which increases the age of the universe by changing
the behavior of the scale factor with time will be subject to
various cosmological tests which are sensitive to that relationship;
these are conventionally used to place limits on the cosmological
constant \cite{cpt}.  Currently the most promising such tests
are direct measurements of deviation from the linear Hubble law
using high-redshift Type Ia supernovae \cite{sne}, and volume/redshift
tests provided by the frequency of gravitational lensing of distant
quasars by intervening galaxies \cite{lenses}.  These have recently
been applied to a number of models with novel dependences of
the scale factor on time, very similar to the scenario discussed
in this paper.  The results to date \cite{limits} seem to indicate that
these tests do not rule out the kind of models considered here,
but may be able to do so in the near future when more data is
available.

Our investigation has been exclusively in the context of an
unperturbed Robertson-Walker cosmology.  The next step is to 
introduce perturbations and discuss CMB anisotropies and the
formation of structure; work in this direction is in progress.
However, it is worth noting some important features of the problem.
There are two powerful effects which distinguish the growth of
perturbations in a VDM cosmology from conventional cold dark matter,
and they tend to affect the power spectrum in opposite ways.  The
first effect is the effectively negative pressure of the coupled 
system.  At zero temperature, perturbations in
vamps grow more rapidly than those in CDM; indeed, perturbations
tend to grow even in the absence of gravity.  The other effect,
meanwhile, is the free streaming of the $\psi$ particles.  The $\psi$'s
decouple while relativistic, and in some respects act as hot dark
matter.  They will tend to flow out of overdense regions, damping
the growth of perturbations until sufficiently late times.  An
accurate appraisal of the magnitude of this process requires 
numerical integration of the evolution equations, as the Boltzmann
equation does not simplify as it would for massless or completely
nonrelativistic particles \cite{mb}.  These two competing effects
are not the entire story; for example, if the $\psi$'s are fermions they
will be prevented from clustering on very small scales by the exclusion
principle \cite{tg}.  The final perturbation spectrum is therefore the 
result of a number of processes, and cannot be reliably estimated
analytically.  In addition, of course, the simple model we
have investigated here may be modified, either by altering the
form of the potential (\ref{potential}) or by introducing other
forms of energy in addition to baryons and vamps ({\it e.g.},
ordinary hot or cold dark matter).

Another direction currently under investigation is the construction
of particle physics models in which vamps may arise.  A possible
origin for the scalar $\phi$ is as one of the moduli of string
theory; our understanding of the nonperturbative effects which
give potentials to such fields is not sufficiently developed to
attempt realistic model building at this time.  In supersymmetric
gauge theories, however, there are (perturbatively) flat directions
whose dynamics are somewhat better understood, and in that
context the search for a model may be more hopeful.  In such a 
scenario there are a number of potentially dangerous effects which
must be avoided; for example, if the expectation value of $\phi$
breaks supersymmetry, it may lead to gradual variations in the parameters
of the standard model as the universe expands.  Such variations are
tightly constrained by a variety of data \cite{constants}.

\section*{Acknowledgments}

We would like to thank Edmund Bertschinger, Edward Farhi, and 
Mark Srednicki for helpful conversations.
This work was supported in part by the National Science Foundation
under grant PHY/94-07195.

\end{document}